\documentclass[%
 aip,
 amsmath,amssymb,
preprint,%
]{revtex4-2}
\usepackage {CJK}
\DeclareSymbolFont{matha}{OML}{txmi}{m}{it}% txfonts
\DeclareMathSymbol{\varv}{\mathord}{matha}{118}

\usepackage{graphicx}% Include figure files
\usepackage{dcolumn}% Align table columns on decimal point
\usepackage{bm}% bold math
\usepackage[utf8]{inputenc}
\usepackage[T1]{fontenc}
\usepackage{mathptmx}
\usepackage{etoolbox}
\usepackage[normalem]{ulem}

\usepackage[T1]{fontenc}
\usepackage[utf8]{inputenc}
\usepackage[absolute,overlay]{textpos}
\usepackage{hyperref}
\usepackage{graphicx}
\usepackage{dcolumn}
\usepackage{bm}
\usepackage{xcolor}
\usepackage{siunitx}
\usepackage[all]{nowidow}
\newcommand{\blu}[1]{{\color{black}{#1}}}

\newcommand*{\citen}{}% generate error, if `\citen` is already in use
\DeclareRobustCommand*{\citen}[1]{%
  \begingroup
    \romannumeral-`\x % remove space at the beginning of \setcitestyle
    \setcitestyle{numbers}%
    \cite{#1}%
  \endgroup
}

\makeatletter
\def\@email#1#2{%
 \endgroup
 \patchcmd{\titleblock@produce}
  {\frontmatter@RRAPformat}
  {\frontmatter@RRAPformat{\produce@RRAP{*#1\href{mailto:#2}{#2}}}\frontmatter@RRAPformat}
  {}{}
}%
\makeatother

\begin{document}
%\begin{CJK*}{GB}{gbsn}
\preprint{AIP/123-QED}

\title[]{Coalescence of Sessile Aqueous Droplets Laden with Surfactant}
% Force line breaks with \\
\author{Soheil Arbabi*}
\email{arbabi@ifpan.edu.pl}
\affiliation{Institute of Physics, Polish Academy of Sciences, Al. Lotnik\'ow 32/46, 02-668 Warsaw, Poland}
\author{Piotr Deuar}%
% \email{deuar@ifpan.edu.pl}
\affiliation{Institute of Physics, Polish Academy of Sciences, Al. Lotnik\'ow 32/46, 02-668 Warsaw, Poland}
\author{Rachid Bennacer}
\affiliation{Universit\'e Paris-Saclay, ENS Paris-Saclay, CNRS, LMPS, 4 Av. des Sciences, 91190 Gif-sur-Yvette, France}
\author{Zhizhao Che}
% \homepage{http://www.Second.institution.edu/~Charlie.Author.}%
\affiliation{%
State Key Laboratory of Engines, Tianjin University, 300350 Tianjin, China%\\This line break forced% with \\
}%
\author{Panagiotis E. Theodorakis*}%
 \email{panos@ifpan.edu.pl}
\affiliation{Institute of Physics, Polish Academy of Sciences, Al. Lotnik\'ow 32/46, 02-668 Warsaw, Poland}

\date{\today}% It is always \today, today,
             %  but any date may be explicitly specified

\begin{abstract}
With most of the focus to date having been on the coalescence of freely suspended droplets, much less is known about the coalescence of sessile droplets, especially in the case of droplets laden with surfactant. Here, we employ large-scale molecular dynamics simulations to investigate this phenomenon on substrates with different wettability. In particular, we unravel the mass transport mechanism of surfactant during coalescence, thus explaining the key mechanisms present in the process. Close similarities are found between the coalescence of sessile droplets with equilibrium contact angles above 90$^\circ$ and that of freely suspended droplets, being practically the same when the contact angle of the sessile droplets is above \blu{140}$^\circ$. Here, the initial contact point is an area that creates an initial contact film of surfactant that proceeds to break into engulfed aggregates. A major change in the physics appears below the 90$^\circ$ contact angle, when the initial contact point becomes small and line-like, strongly affecting many aspects of the process and allowing water to take part in the coalescence from the beginning. We find growth exponents consistent with a {2}/{3} power law on strongly wettable substrates but no evidence of linear growth. Overall bridge growth speed increases with wettability for all surfactant concentrations, but the speeding up effect becomes weaker as surfactant concentration grows, along with a general slowdown of the coalescence compared to pure water. Concurrently the duration of the initial thermally limited regime increases strongly by almost an order of magnitude for strongly wettable substrates.
\end{abstract}
\maketitle
%\end{CJK*}
\section{Introduction}
Many natural phenomena involve the coalescence of fluid droplets.\cite{Paulsen2014,yoon2007coalescence,khodabocus2018scaling,perumanath2019droplet,eggers1999coalescence,aarts2005hydrodynamics,Sprittles2012,dudek2020,Rahman2019,Berry2017,Somwanshi2018,Kirar2020,Bayani2018,Brik2021,Anthony2020,Kern2022,Heinen2022,Geri2017,Abouelsoud2021,Hilaire2023}
For example, this is a fundamental process 
that determines the distribution and coalescence rate of raindrops
in atmospheric aerosols.\cite{Bowen1950,denys2022lagrangian,Berry1974,kovetz1969effect}
Apart from such natural processes, droplet coalescence is relevant for 
many industrial applications as well, such as inkjet printing,\cite{Singh2010}
microfluidics,\cite{feng2015advances,dudek2019microfluidic,Choi2012,Fair2007}
and water treatment during crude oil and natural gas 
separation.\cite{thomas2019led,Wijshoff2010}
Further control of the process may involve the use of various additives,\cite{Dekker2022,Calvo2019,Arbabi2023,Varma2021,Sivasankar2021,Sivasankar2022,Otazo2019,Vannozzi2019} 
such as surfactant,\cite{Politova2017,Weheliye2017,Soligo2019,Dong2017,Dong2019,Kovalchuk2019,Botti2022,Amores2021,Kasmaee2018,nowak2017bulk,nowak2016effect,jaensson2018tensiometry,narayan2020zooming,ivanov1999flocculation,tcholakova2004role,langevin2019coalescence,Lu2012,Velev1993,Suja2018}
which can reduce surface tension at fluid interfaces, crucial in multi-phase systems.
For example, surfactant can stabilize droplets' surface
or prevent their coalescence, thus improving the
bio-compatibility in certain systems\cite{baret2012surfactants}
or affecting the fusion, mixing, and manipulation of droplets
in microfluidic devices.\cite{elvira2022materials} 
While many efforts have thus far been taken to
understand coalescence phenomena at a fundamental level,\cite{Paulsen2014,
khodabocus2018scaling,hopper1993coalescence}
there are still numerous aspects of this process that
require further investigations, such as the coalescence of sessile droplets with surfactant.

In experiments,\cite{bruning2018delayed,yi2022cascade,schwarzwalder2023experimental,sun2020experimental,
lee2012coalescence,chen2020understanding,park2021coalescence}  
droplet coalescence has been studied for various
conditions (\textit{e.g.}, in the presence of applied electric fields) 
and geometries (\textit{e.g.}, micro-channels, fibers, \textit{etc.}),
mainly by means of high-speed imaging and electrical sensing.
Due to resolution limitations, the focus of these investigations
has mainly been placed on unveiling macroscopic properties of
the process.
\cite{nowak2016effect, narayan2020zooming, nowak2017bulk, duchemin2003inviscid, leal2004flow,Chinaud2016}
For example, the effect of surfactant concentration
on droplet coalescence has been investigated by 
high-speed imaging.\cite{nowak2016effect}
When an asymmetry in surfactant concentration of the coalescing droplets is present, 
Marangoni flow was observed and the curvature on either side of the growing 
bridge was different. Numerical continuum approaches have also been employed to
complement understanding of droplet coalescence.
\cite{Soligo2019,shaikh2021numerical,gu2022numerical,yeo2003film,hu2000drop,mansouri2014numerical,eggers1999coalescence,duchemin2003inviscid,Sprittles2012,Heinen2022,Chen2021,Anthony2020,leong2020droplet}
However, such methods suffer from inadequate resolution in
capturing the mass transport mechanism of surfactant during
coalescence or resolving the initial contact of the droplets.
The singularity at the contact point still remains a challenge for continuum simulation.\cite{Sprittles2012}
On the other hand, molecular dynamics (MD) simulation can naturally resolve
the contact region at the molecular level, to observe the start of the
coalescence process.

\begin{figure}[bt!]
\includegraphics[width=0.55\textwidth]{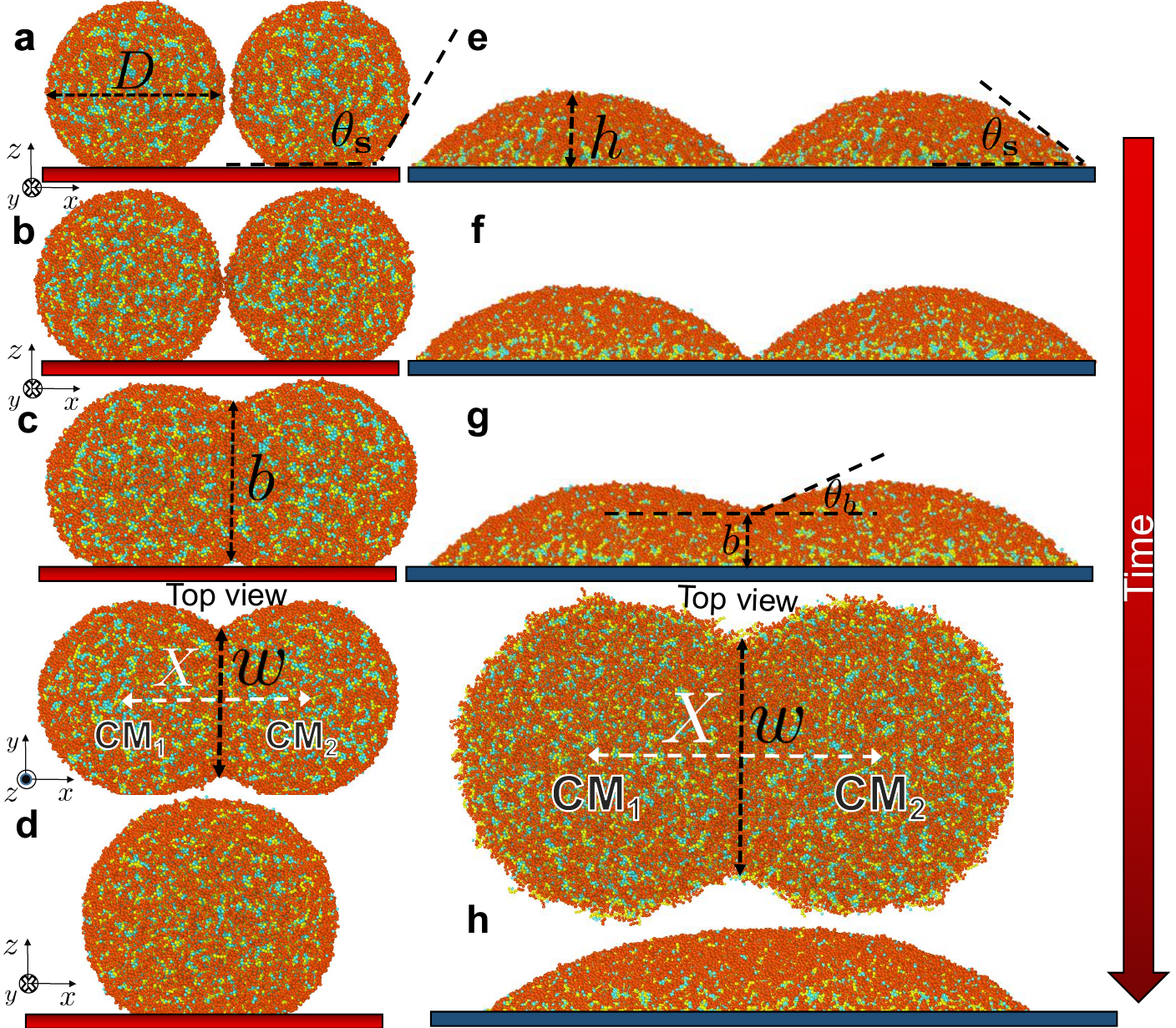}
\caption{\label{fig:1}
Different stages of coalescence of surfactant-laden droplets
on non-wettable (a--d, $\theta_s \simeq \blu{142}^{\circ}$)
and wettable (e--h, $\theta_s \simeq 49^{\circ}$) substrates, with
surfactant concentration above 
CAC (35.48~wt\%). 
$b$ is the bridge height and $w$ the bridge width.
\blu{$X$ is distance between the centers of mass of the two droplets in the $x$-direction},
while $\theta_s$ is the equilibrium contact angle of 
the droplet. $\theta_b$ is the angle formed at the
bridge at the liquid--gas surface. The stages of coalescence shown here are:
(a, e) Initial approach ($t=t_c$);
(b, f) Moment of pinching ($t=t_c$);
(c) Developed bridge ($t=t_c+196.25~\tau$) and 
(d) final, equilibrium state of the system ($t=t_c+1483.75$) in 
the case of the non-wettable substrate.
(g) Developed bridge ($t=t_c+1028.75~\tau$) and
(h) final, equilibrium state of the system ($t=t_c+3161.25~\tau$)
in the case of the wettable substrate.
The snapshots of the system were obtained using  Ovito software.\cite{Stukowski2010} }
\end{figure}

\begin{figure}[bt!]
\includegraphics[width=0.25\textwidth]{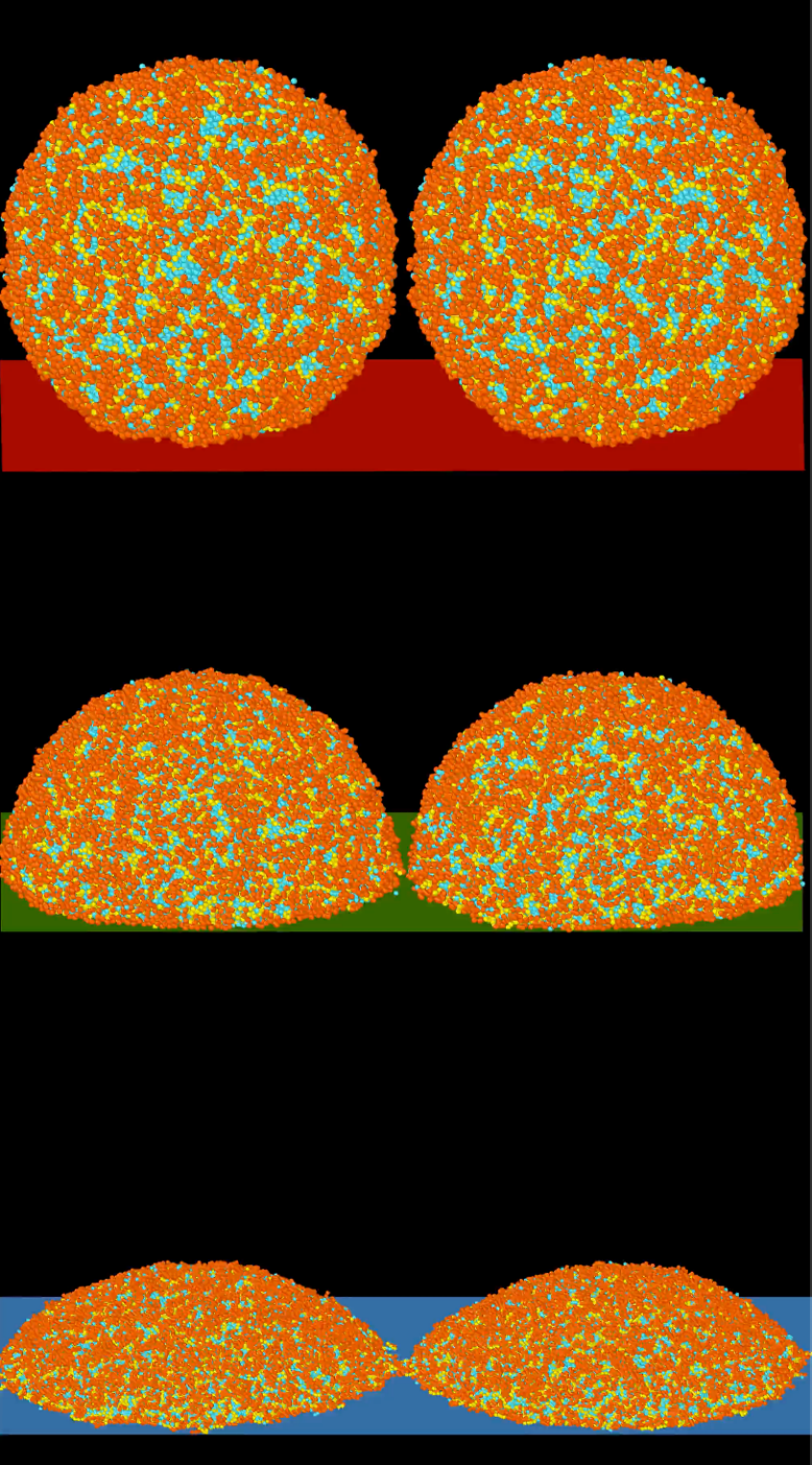}
\caption{\label{fig:figMovie1} (Multimedia view) Coalescence of droplets with surfactant concentration (35.48~wt\%) on substrates with different wettability (upper row, non-wettable; middle row, intermediate; lower row, wettable). 
}
\end{figure}

In general, droplet coalescence is a non-equilibrium process
that occurs in three stages. 
The first stage involves the initial approach of droplets, 
when they come close enough to interact (pinching/contact) and form the so-called 
bridge (Figure~\ref{fig:1}a, e and pinching b, f). 
This is followed by a bridge-growth stage
that leads to the reshaping (third stage) of the droplets (Figure~\ref{fig:1}c, g)
and eventually the single spherical-cap shape for sessile droplets,
which is the final equilibrium and minimum energy state of the system (Figure~\ref{fig:1}d,h). The time evolution of this process on substrates with different wettability is illustrated in Figure~\ref{fig:figMovie1} (Multimedia view).
The coalescence process is governed by the interplay of viscous, inertial and surface-tension
forces, as the system tries to decrease the surface tension.\cite{verdier2002understanding}
In particular, from the perspective of fluid dynamics,
viscous forces are expected to play an important role
in the initial bridge growth, which is driven primarily by
molecular interactions. At a later stage of the
bridge growth, inertial effects are expected to take over. 
In the case of freely suspended droplets,  a linear scaling of the bridge radius, $b$, 
in time ($b \propto t$) and  logarithmic corrections $b\propto t \ln t$ have been suggested 
for the viscous regime.\cite{hernandez2012symmetric,aarts2005hydrodynamics}
For the inertial regime, a power law scaling ($b\propto t^{1/2}$) has been 
proposed.\cite{eggers1999coalescence} However, this classification is still under debate 
in the literature, and an inertially limited viscous (ILV) regime has also been 
reported.\cite{paulsen2013approach,Paulsen2014} Moreover, an initial thermal regime, 
which is inaccessible to experiments has been identified by means of 
all-atom\cite{perumanath2019droplet} and coarse-grained (CG)\cite{Arbabi2023b,Arbabi2023c} 
MD simulation. This regime arises due to collective thermal motion of particles
at the droplets' surface. Here, all-atom simulation has proposed a scaling law for a 
length scale characterizing the extent of the thermal fluctuations, namely
$l_T  \approx {\left ( k_BT/\gamma \right) }^{1/4} R^{/1/2}$,
where $k_B$ is Boltzmann's constant, $T$ the temperature,
$\gamma$ the surface tension, and $R$ the droplet radius.\cite{perumanath2019droplet}
When the bridge radius is smaller than this length
scale  ($b<{l_T}$), the bridge grows mainly due to 
thermal motion of particles, while later ($b>{l_T}$) 
hydrodynamic effects are expected to dominate.
Recent MD simulations of a CG model have indicated
the presence of the thermal regime and subsequent
inertial scaling law ($b\propto t^{1/2}$) for the bridge
growth for both aqueous droplets with and 
without surfactant.\cite{Arbabi2023b,Arbabi2023c}

In the case of the coalescence of sessile droplets,\cite{Tao2023,hernandez2012symmetric,eddi2013influence,ristenpart2006coalescence,
narhe2008dynamic,
kapur2007morphology,Kovalchuk2019,bazazi2020retarding,yatsyshin_kalliadasis_2021,Arbabi2023}
for water droplets on non-wettable substrates
(contact angle $\theta_s \geq 90^{\circ}$, Figure~\ref{fig:1}) 
it has been suggested that the bridge grows with time as $b \propto t^{1/2}$, 
as has been observed in the case of freely suspended
droplets. In contrast, for wettable substrates, 
namely $\theta_s < 90^{\circ}$, the bridge is predicted to
grow with a new power law $b \propto t^{2/3}$.\cite{eddi2013influence}
Moreover, experiments of droplets with an equilibrium 
contact angle in the range $10^{\circ}-56^{\circ}$
suggest that the bridge height roughly grows in time with power-law exponents
between 0.50 and 0.86. Data was seen to follow the scaling law
for the entire range of time in the case of $10^\circ$ contact angle, while the rest of
the cases studied deviated from this law at longer times.\cite{lee2012coalescence}
The bulk fluid properties also affect the bridge growth. 
For example, in the case of polymer droplets\cite{Arbabi2023,varma2020universality,varma2022elasticity}, their viscosity can result in 
a lower rate of coalescence in comparison with
water droplets. One finds power-law exponents lower than 1/2 for nonwettable 
substrates (equilibrium contact angle greater than $90^\circ$)
and power-law exponents lower than 2/3 for wettable substrates.\cite{Arbabi2023}
The bridge width, $w$ (Figure~\ref{fig:1}), grows
as $w \propto t^{0.5}$.\cite{ristenpart2006coalescence,lee2012coalescence,narhe2008dynamic} 
However, a linear scaling $w \propto t$  has also been suggested
for droplets of high viscosity.\cite{hernandez2012symmetric} 
To our knowledge, the regimes of applicability of the above findings have not been 
investigated in experiments nor in  modeling that would include the early-time 
molecular level physics.

In view of the many unknowns in the coalescence of sessile surfactant-laden droplets, 
we embarked on investigating this phenomenon
by using large-scale molecular dynamics simulations
of a coarse-grained force-field, considering a comprehensive range of possible scenarios. 
Hence, our study includes a range of surfactant concentrations below and
above the critical aggregation concentration (CAC) and
substrates with different wettability, both wettable ($\theta_s<90^\circ$)
non-wettable ($\theta_s>90^\circ$), and those with
equilibrium contact angle of about $90^\circ$. 
In all these cases, we have explored the mass transport
mechanism of surfactant, which provides insights into
the details of the coalescence process, analyzed the dynamics
of the bridge growth, which characterizes the rate of
coalescence, as well as \blu{studied} the bridge angle and the velocity 
of approach.\cite{verdier2000coalescence}
It turns out that while the coalescence of sessile surfactant-laden droplets
on non-wettable substrates shares similarities with the coalescence
of freely suspended droplets, significant differences in the
mass transport mechanism and rate of coalescence appear
for wettable substrates. The following section presents our 
MD simulation model and methods.
Our results are discussed in Section~\ref{results}.
Finally, the conclusions and possible suggestions 
for follow-up research are discussed in Section~\ref{conclusions}.

\section{Simulation model and methods}
\label{model}
Our system consists initially of two droplets placed close to each other as shown in 
Figure~\ref{fig:1}a,e, that is a distance below the cutoff range of the
interactions between beads in order to initiate the coalescence of the droplets.
We have considered concentrations above and below the CAC for
C10E4 surfactants.
The details of the model for the interactions between the 
different components and the validation for various 
surfactants are taken the same here as used in a number of previous related
studies,\cite{Theodorakis2015Langmuir,theodorakis2015modelling,Theodorakis2019,Lobanova2015,lobanova2014development, lobanova2016saft, morgado2016saft}
and were obtained through the Mie-$\gamma$ Statistical Associating Fluid Theory (SAFT Mie-$\gamma$).\cite{Muller2014,Lobanova2015,Lafitte2013,Avendano2011,Avendano2013}
The MD simulations were carried out in the
canonical ensemble with the Nos\'e--Hoover
thermostat as implemented in LAMMPS software.\cite{Plimpton1995,LAMMPS}
After equilibration of a single droplet on the specific substrate, 
it was cloned to produce the second droplet and the surrounding vapor for the start of 
the coalescence \textit{in silico} experiment (Figures~\ref{fig:1}\blu{a,e}).

The force-field has been validated for a range of
water--surfactant systems with a focus on reproducing
key properties, such as surface tension and phase
behavior.\cite{theodorakis2015modelling, Theodorakis2015Langmuir, Theodorakis2019, theodorakis2019molecular, Theodorakis2014, lobanova2014development, lobanova2016saft, morgado2016saft}
In particular, interactions between the various
CG beads representing different chemical units of the system
are described via the Mie potential %reading
\begin{equation}
\label{equation_mie}
    {U(r_{\rm ij})} = C\epsilon_{\rm ij} \left[ \left({\frac{\sigma_{\rm ij}}{r_{\rm ij}}}\right)^{\lambda_{\rm ij}^{\rm r}} - \left({\frac{\sigma_{\rm ij}}{r_{\rm ij}}}\right)^{\lambda_{\rm ij}^{\rm a}}\right], 
    \; {\rm for} \; r_{\rm ij} \leq r_{\rm c},\\
    \end{equation}
    where
\begin{equation*}
    C = \left(\frac{\lambda_{\rm ij}^{\rm r}}{\lambda_{\rm ij}^{\rm r} - \lambda_{\rm ij}^{\rm a}}\right){\left( \frac{\lambda_{\rm ij}^{\rm r}}{\lambda_{\rm ij}^{\rm a}}\right)}^{\frac{\lambda_{\rm ij}^{\rm a}}{\lambda_{\rm ij}^{\rm r} - \lambda_{\rm ij}^{\rm a}}},
\end{equation*}
%Here, 
and i and j represent the bead types. Hence, $\sigma_{\rm ij}$ is an
effective size of these beads, while $\epsilon_{\rm ij}$ sets the interaction 
strength between beads of type i and j. One takes $\lambda_{\rm ij}^a=6$,
which is connected to representing the
dispersive interactions between the different particles, 
while $\lambda_{\rm ij}^r$ serves as a fitting parameter
in the SAFT model and can vary. Finally, $r_{\rm ij}$ 
is the distance between any pair of beads, which 
interact when their distance is below a cutoff value set to
$r_c = 4.583$~$\sigma$, where $\sigma$ is the overall unit of length.
The rest of the units are $\epsilon$ for the energy, 
$m$ for the mass, while $\tau$ is the natural 
MD time unit $\tau = \sigma{(m/{\epsilon})}^{0.5}$.
In real units, we consider the simulation to correspond to: 
$\sigma = 0.43635$~nm, $\epsilon / k_B = 492$~K, $m=44.0521$~amu
and $\tau = 1.4062$~ps.
The integration of the equations of motion was carried out with an
integration time-step $\delta t = 0.005$~$\tau$,
while the temperature was set to $k_BT/\epsilon = 0.6057$, 
which would correspond to $T=25$~$^\circ$C in real units.

\begin{figure}[bt!]
\centering
\includegraphics[width=0.55\columnwidth]{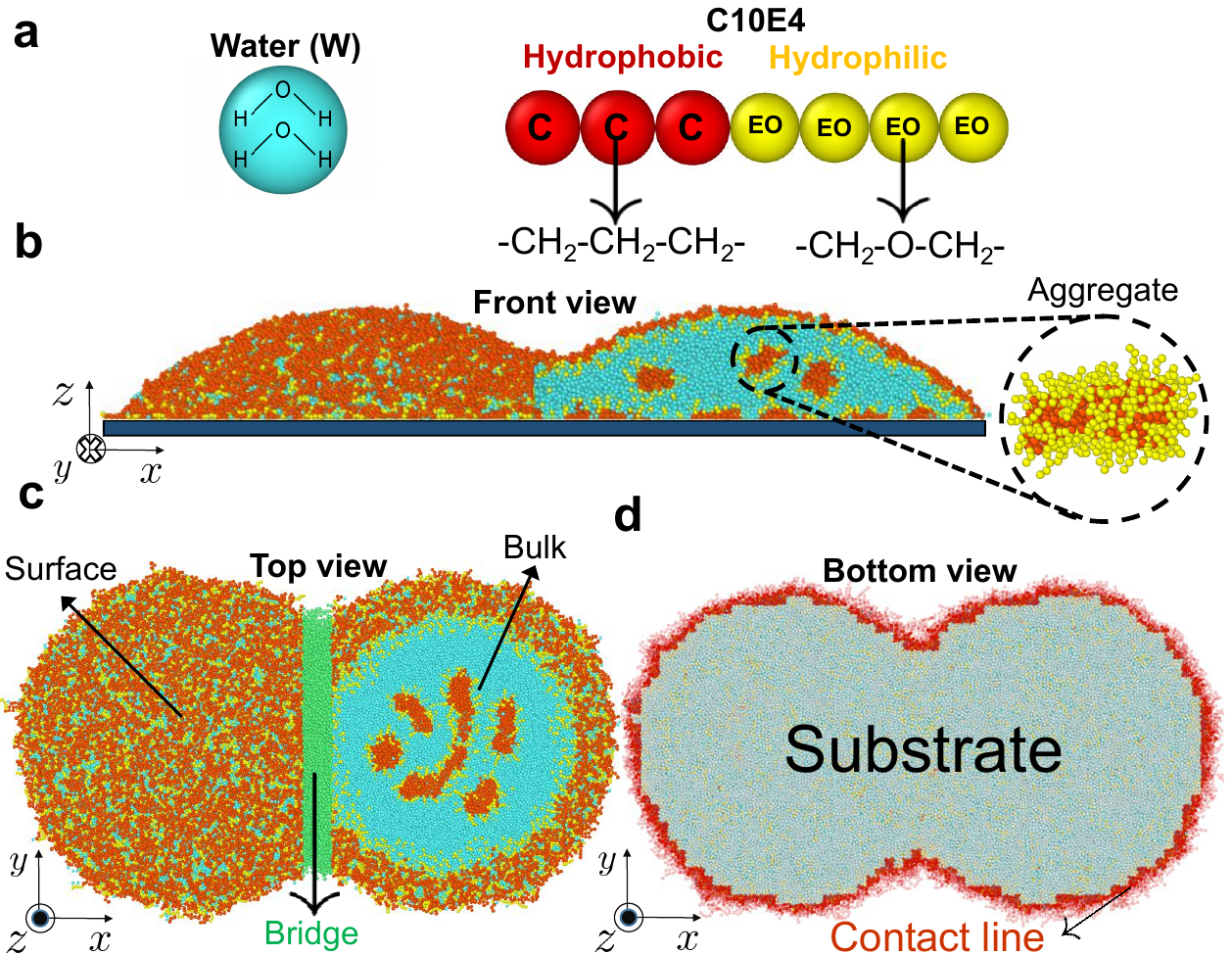}
\caption{\label{fig:2}
\blu{Description} of model and notation.
(a) Coarse-grained representation of water and C10E4 surfactant.
(b\blu{--d}) External and cross-section views are shown to highlight the bulk, interfaces, and
contact line of the droplet ($\theta_s =49^{\circ}$, with concentration above CAC). 
Surrounding water vapor is omitted for the sake of clarity.
(b) Front view; (c) Top view; (d) Bottom view.
}
\end{figure}

Here, we have chosen for our investigations
the C10E4 surfactant, 
which has also been previously studied in the context of 
the coalescence of freely suspended droplets\cite{Arbabi2023b,Arbabi2023c} and
superspreading.\cite{Theodorakis2015Langmuir,theodorakis2015modelling,Theodorakis2019}
The above studies found other CmEn type surfactants to give similar
behavior,\cite{Arbabi2023b,Arbabi2023c} 
so here we only consider C10E4 as representative of the whole family.
Then, the CG representation of the C10E4 (Figure~\ref{fig:2}) uses
hydrophobic alkane CG `C' beads with each one representing a $\rm -CH_2-CH_2-CH_2-$ 
group of atoms. Hydrophilic CG `EO' beads represent
oxyethylene groups $\rm -CH_2-O-CH_2$. Finally, water CG `W' beads
correspond to two water molecules. The values of the nonbonded interaction
parameters between CG beads are reported in \blu{Table~S1 of the Supplementary Material (SM) }
and their masses in \blu{Table~S2 of the SM}.

To link the alkane and oxyethylene beads together,
each consecutive pair of beads along the surfactant chain interacts
via a harmonic potential of the form
\begin{equation}
\label{equation_bonded1}
    V(r_{\rm ij}) = 0.5k(r_{\rm ij}-\sigma_{\rm ij})^2
\end{equation}
where  $k = 295.33$~$\epsilon/\sigma^2$.
In addition, each consecutive triad of EO beads along the chain experiences a
harmonic angle potential, \textit{i.e.}
\begin{equation}
\label{equation_bonded2}
    V_\theta(\theta_{\rm ijk}) = 0.5k_\theta(\theta_{\rm ijk}- \theta_0)^2 
\end{equation}
where $\theta_{\rm ijk}$ is the angle between consecutive
beads i, j=i+1 and k=j+1. Constants are
$k_\theta = 4.32$~$\epsilon/$rad$^2$, and $\theta_0 = 2.75$~rad for
the equilibrium angle.

The total number of beads per initial droplet in each simulation
was $10^5$ %beads 
and two different surfactant concentrations were
considered, namely 6.25~wt\%, which is below CAC, and 35.48~wt\%,
which is well above, with CAC roughly being 7.5~wt\%.\cite{Arbabi2023b,Arbabi2023c}
The latter concentration is taken as an average, since the number
of water molecules in the liquid phase (droplet) fluctuates. 
The wettability of the smooth, unstructured substrate
was also tuned, as described below, covering
a range of wettable ($\theta_s<90^\circ$) and non-wettable ($\theta_s>90^\circ$) types of 
substrate (Figure~\ref{fig:1}). The exact values of the equilibrium contact angle of
single droplets are reported
in \blu{Table~S3 of the SM} along with the corresponding droplet dimensions \blu{(Table~S4 of the SM)}.
Finally, to determine the beads
belonging to the liquid phase (droplet), a cluster analysis has been 
performed.\cite{Stillinger1963,Theodorakis2011}

To vary the wettability of the substrate, we need to
define the interactions between the droplet beads and the substrate.
This can be done by using the
combining rules defined in SAFT$\gamma$-Mie\cite{Lafitte2013}:
\begin{equation}
\label{combining_rule_sigma}
    \sigma_{\rm ij} = \frac{\sigma_{\rm ii}+\sigma_{\rm jj}}{2},
\end{equation}

\begin{equation}
\label{combining_rule_epsilon}
    \varepsilon_{\rm ij} = \frac{\sqrt{\sigma_{\rm ii}^3\sigma_{\rm jj}^3}}{\sigma_{\rm \blu{ij}}^3}\sqrt{\varepsilon_{\rm ii}\varepsilon_{\rm jj}}.
\end{equation}
and start from
the defined interactions for the liquid phases as given in \blu{Table~S1 of the SM.}
By tuning the cross interaction $\varepsilon_{ws}$ (``s'' indicates the substrate), 
the empirical relationship between $\varepsilon_{ws}$ and the contact angle  of 
pure water droplets \blu{(Figure~S1 of the SM)} can be found, \blu{and is} %as 
shown in \blu{Figure~S2} of the SM. Then, by using the above combining rules,
the parameter $\varepsilon_{ss}$ corresponding to a given $\varepsilon_{ws}$
can be obtained.
We shall note here that the exact value of the $\sigma_{ss}$ parameter
is not important for our studies and is set to unity for simplicity. 
Similarly, $\lambda_{sj}^r=9$ \blu{is set} for all interactions involving the substrate.
Based on the \blu{k}nowledge of the latter parameters and  the interactions
$\sigma_{ww}$ and $\varepsilon_{ww}$ (\blu{Table~S1 of the SM}),
the interaction of the substrate with the
surfactant beads can be obtained as well, again 
using Equations~\ref{combining_rule_sigma}-\ref{combining_rule_epsilon}. 
After obtaining all the interaction parameters,
equilibrium simulations are run and the surfactant-laden droplet's contact angle is determined
by using the method of Ref.~\citen{theodorakis2015modelling} \blu{(see also text and Figure~S1 of the SM)}
with data reported in \blu{Table~S3 of the SM}.  
\blu{It might be argued however that estimating the angles can in general be sensitive to the details of the definition of a sharp interface, as well as to the fitting procedure.\cite{yatsyshin_kalliadasis_2021,Mugele2002} Also, models that could take into account the disjoining pressure effects, very relevant in the context of droplets on solids substrates, might perform better than fitting spherical caps to nanodroplets.\cite{yatsyshin_kalliadasis_2021}}
Moreover, according to a previous study,\cite{theodorakis2015modelling} the size of the droplet is 
large enough to guarantee that the equilibrium contact angle
does not depend on the size of the droplet, which makes the interaction values
valid for both larger and smaller droplets.
\blu{Table~S3 of the SM} provides the exact values for all of the cross interactions
between the beads and the substrate used in the simulations.
Our choice here covers the relevant parameter space 
for surfactant concentration (above and below CAC)
and substrate types (wettable, non-wettable, and about $90^\circ$). 

The mass transport of surfactant molecules is investigated by
tracking the motion of each individual molecule between
the various parts of the droplet, which are illustrated
in Figures~\ref{fig:2}c-d. These different parts are the 
liquid--gas (LG) surface for the left and right droplets (the center
of the coordinate system is taken to be in the middle between
the two droplets where the bridge forms), the bulk
of the left and right droplets, the solid--liquid (SL)
interface for the left and right droplets, 
the contact line for the left and right droplets,
and the LG and SL interfaces, bulk, and contact line
of the bridge as shown in Figure~\ref{fig:2}.
Hence, this makes a total of twelve different regions
where each surfactant molecule can belong. 
Further discussion and details on the calculation
of the probabilities related to the motion of 
surfactant between these regions describing the mass
transport mechanism of surfactants can be found in the SM.
Finally, we have calculated the density profiles
of the water and surfactant during coalescence,
and the approach distance and velocity of the
droplets\cite{Arbabi2023}, which will be further
discussed in the following section.

\begin{figure}[b!]
\centering
\includegraphics[width=0.45\columnwidth]{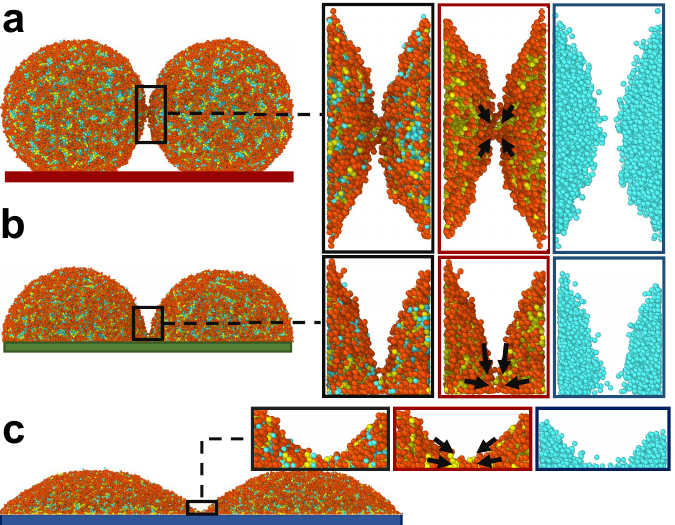}
\caption{\label{fig:3}
The pinching moment ($t \approx t_c$) in the
case of (a) a non-wettable substrate ($\theta_s>90^\circ$),
(b) an intermediate case ($\theta_s \approx 90^\circ$), and
(c) a wettable substrate ($\theta_s<90^\circ$).
Water participation in the coalescence process is
more pronounced in the case $\theta_s < 90^\circ$,
while the cases $\theta_s \geq 90^\circ$ behave
very similarly to
freely suspended droplets.\cite{Arbabi2023b,Arbabi2023c}
\blu{Snapshots were obtained using Ovito
software.\cite{Stukowski2010} }}
\end{figure}

\section{Results and discussion}
\label{results}
\subsection{Surfactant mass transport mechanism}
Our previous studies\cite{Arbabi2023b,Arbabi2023c} of the coalescence of freely
suspended droplets have shown that surfactant
plays an ever increasing role at the pinching point
as its concentration in the droplets increases,
while water has a smaller effect in initiating the coalescence 
process. Similar behavior is also observed in the case of $\theta_s>90^\circ$ 
shown in Figure~\ref{fig:3} for sessile droplets,
since the bridge formation starts far from the substrate (Figure~\ref{fig:3}a).
For this reason, the bridge region initially is not affected by the
presence of the substrate and the bridge angle is very steep
($\theta_b\approx80^\circ$, for example, see \blu{Figure~S6} of the SM)
when pinching occurs.
For the same reason, we find that the growth of the bridge occurs symmetrically in both
the $y$ and $z$ directions ($b$, $w$, Figure~\ref{fig:1}). 
In contrast, in the case of wettable substrates, water molecules
participate in the pinching process from its onset \blu{(Figure~\ref{fig:3}c)}. 

\begin{figure}[bt!]
\centering
\includegraphics[width=\columnwidth]{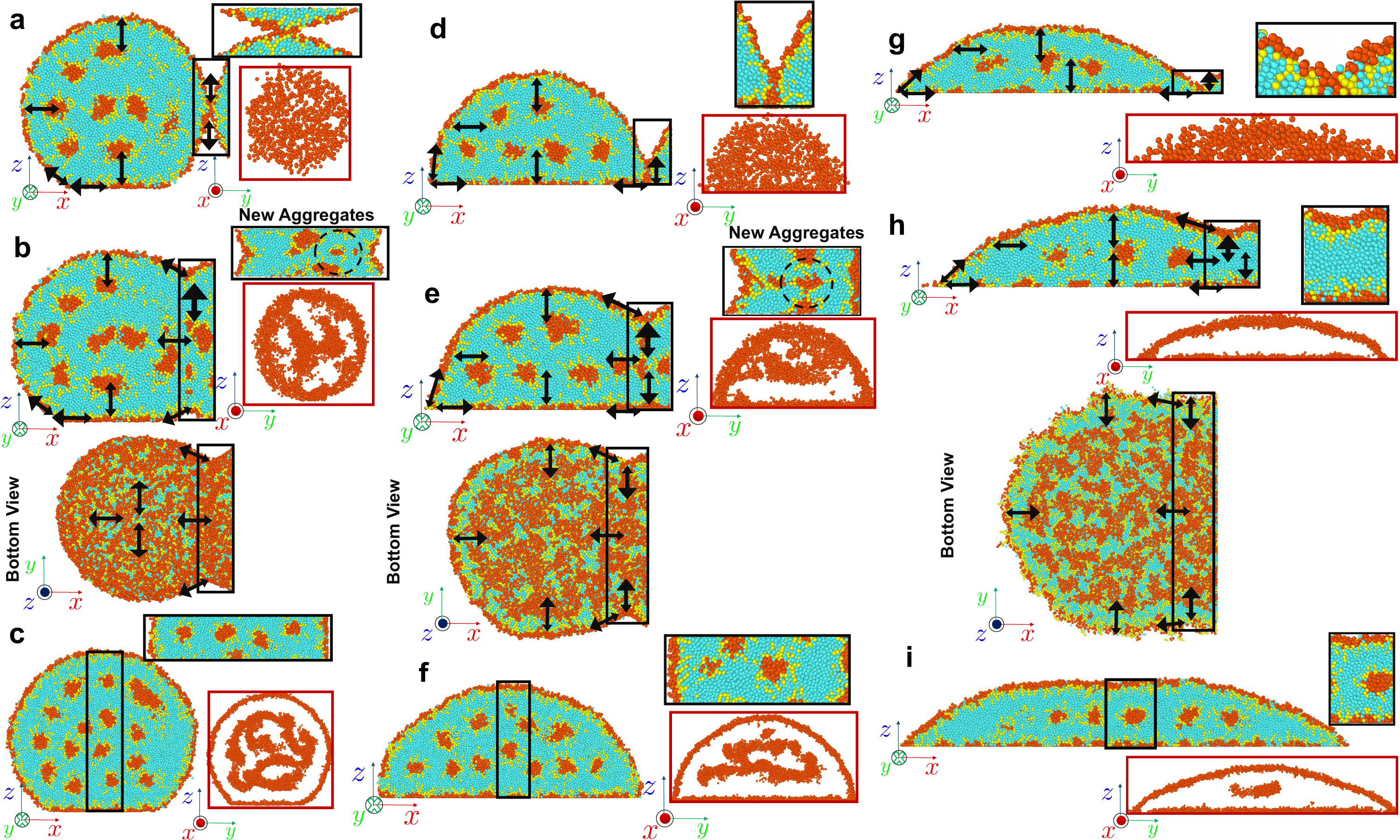}
\caption{\label{fig:4}
Snapshots of coalescing droplets with C10E4 surfactant and surfactant mass 
transport on substrates of different wettability.
The surfactant concentration in the liquid phase is above CAC, namely 35.48~wt\%. 
The size of the arrow heads reflects the probabilities
associated with surfactant transport to the different droplet
areas (see \blu{Table~S5} of the SM for further details). 
(a--c) corresponds to the case 
$\theta_s \simeq \blu{142}^\circ$ with snapshots obtained 
at times (a) $t_{\rm c}+17.5~\tau$,
(b) $t_{\rm c}+185.0~\tau$,  and
(c) $t_{\rm c}+1733.7~\tau$. 
(d--f) $\theta_s \simeq 94^\circ$, with  snapshots shown  
at times (d) $t_{\rm c}+28.75~\tau$, 
(e) $t_{\rm c}+315.0~\tau$, and
(f) $t_{\rm c}+1415.0~\tau$.
(g--i) $\theta_s \simeq 49^\circ$ with snapshots 
at times (g) $t_{\rm c}+200.0~\tau$, 
(h) $t_{\rm c}+1358.7~\tau$, and
(i) $t_{\rm c}+2200.0~\tau$. 
Snapshots in (a), (d), and (g) are soon after the end of the 
thermal regime. Snapshots (b) and (e) illustrate a clearly developed
bridge with new aggregates formed in its bulk or
additional monomers remaining at the bridge region.
(h) shows a clearly developed bridge highlighting
the absence of aggregates in the 
case $\theta_s < 90^\circ$. (c), (f), and (i) correspond to cases 
of a fully developed bridge. Magnified views of the bridge region and 
its cross-section (showing only surfactant hydrophobic beads in the
bridge region, red) are attached above and to the right of the
snapshots, respectively. Snapshots were obtained using Ovito
software.\cite{Stukowski2010} }
\end{figure}

\begin{figure}[bt!]
\centering
\includegraphics[width=\columnwidth]{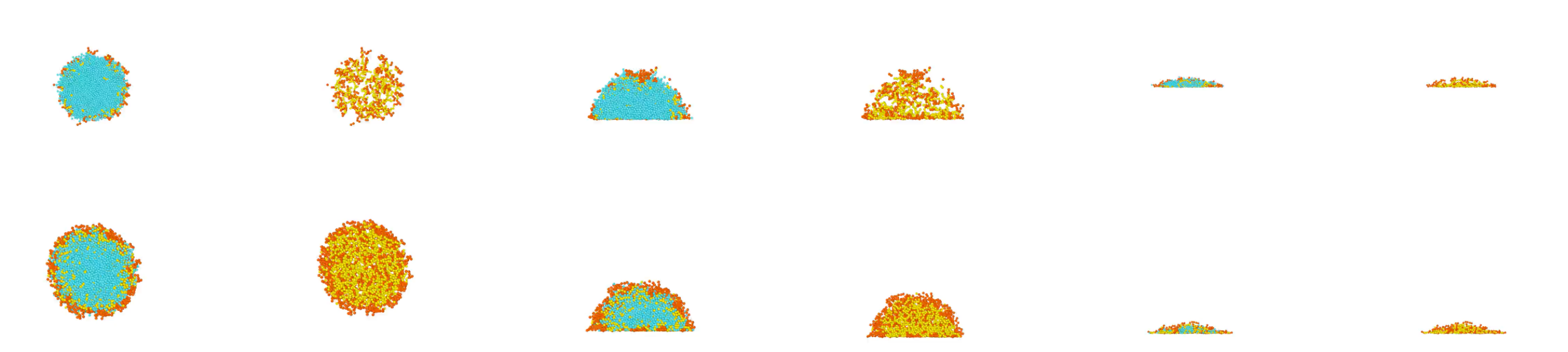}
\caption{\label{fig:figMovie2} (Multimedia view)
Cross-section of the bridge during coalescence on different substrates (first two columns, $\theta_s>90^\circ$, middle two columns, $\theta_s\approx90^\circ$, and last two columns, $\theta_s<90^\circ$). Upper row shows the case of droplets with concentration lower than CAC (6.25~wt\%), while the lower row  with concentration above CAC (35.248~wt\%). All beads (odd columns) or surfactant beads only (even columns) are shown at each cross-section \blu{at the bridge}.   }
\end{figure}

To quantify mass transport, we count the surfactant molecule movements 
between the 12 regions identified in Figure~\ref{fig:2}. The raw numbers
are reported in \blu{Table~S5} of the SM.
These set the intensity of the arrows in Figure~\ref{fig:4}, indicating the dominant
direction of surfactant transport.
Figures~\ref{fig:4}a--c present the dominant surfactant
transport processes for the case $\theta_s>90^\circ$.
As in the case of freely suspended
droplets, an interface film of surfactant initially forms consisting of
surfactant from the LG surfaces of the droplets that
come into contact.
Figure~\ref{fig:figMovie2} (Multimedia view) illustrates this in the case of non-wettable substrates, as well as the absence of this film
when coalescence takes place on wettable substrates.
The perimeter of the bridge is expected to linearly
grow with $b$, while the area of the film
increases as $b^2$. Moreover, the data in \blu{Table~S5} of the SM shows that
the dominant movement of surfactant is towards the LG surface of the droplets, as in
the case of freely suspended droplets. 
Since this movement towards the LG surface occurs and the area of contact
between the droplet grows, 
the surfactant concentration in the film decreases, the film ruptures and 
some surfactant remains in the form of aggregates. 
Due to the lack of space at the LG interface of the merged droplets, 
surfactant from the newly formed aggregates in the bridge bulk  cannot be
accommodated at the LG interface.
Also, we observe surfactant transport away from the bridge from the 
SL interface toward the LG surface through the contact line. 
Other transport processes away from the bridge
are insignificant during coalescence in the case of non-wettable droplets.
When the contact angle $\theta_s \approx 90^\circ$,
the pinching is similar to the case of droplets
with contact angle $\theta_s > 90^\circ$ (Figure~\ref{fig:3})
as well as the case of freely suspended droplets.\cite{Arbabi2023b,Arbabi2023c}
Finally, $\theta_b$ has a value of around $90^\circ$ at the pinching \blu{moment} \blu{(for example, see Figure~S6 of the SM)}.

The case $\theta_s<90^\circ$ shows a somewhat different behavior.
The transport toward the LG surface is higher than in the cases where
$\theta_s \geq 90^\circ$, and unlike the previous cases, 
we do not see the formation of new aggregates
as coalescence takes place [Figure~\ref{fig:figMovie2} (Multimedia view)]. This can be attributed to several causes. 
Firstly, the amount of surfactant at the bridge
is smaller in the case of $\theta_s<90^\circ$,
due to the higher curvature of the droplets. In this case, there is
only a small portion of the droplet surfaces that come into contact,  at 
the contact line of the droplets. In contrast, for non-wettable substrates,
a large portion of the surfaces of the two droplets come into contact forming
a film. This major contact area difference may also explain the higher degree of 
participation of water molecules in the pinching process in the case of
wettable substrates. Secondly, there is much less space in the bridge
to form aggregates from any excess of
surfactant that does not start at
the bridge's LG surface. Finally, we note that surfactant transport from
the contact line towards the SL and LG interface, is overall more
pronounced than in the case of non-wettable substrates. This might be due to
the immediate start of the decrease of the contact-line length 
as the droplets merge, resulting in a greater
excess of surfactant in the contact-line region, and a greater
migration to the LG and SL interfaces.

\begin{figure}[b]
\centering
\includegraphics[width=0.5\columnwidth]{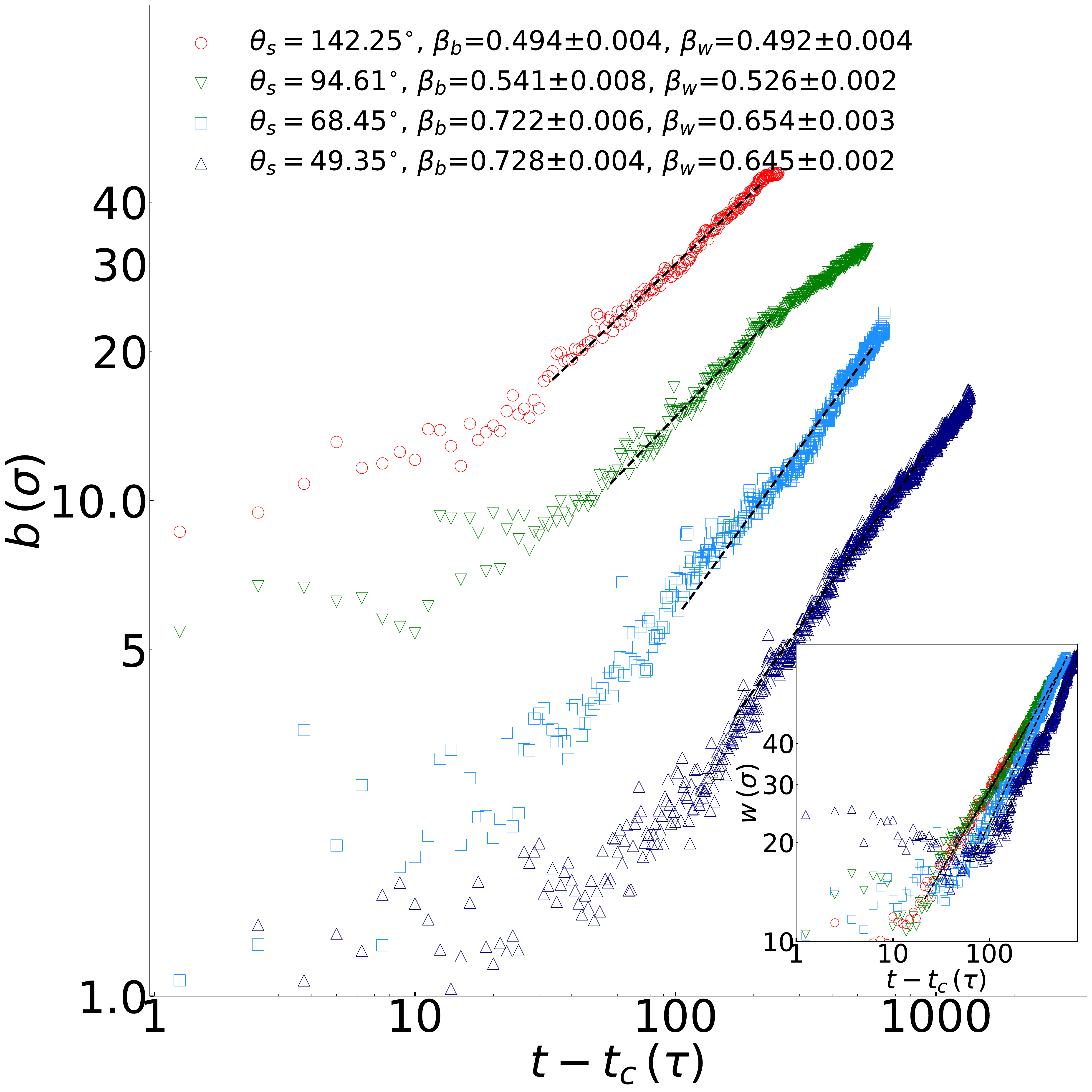}
\caption{\label{fig:5}
Bridge height, $b$, as a function of time, $t$, starting from the
pinching moment, $t_c$. Inset shows the bridge width, $w$. 
Results for \blu{droplets} with surfactant concentration 35.48~wt\% \blu{on substrates with} different 
wettability, as indicated. The values of the power-law exponents for $b$ and
$w$ are $\beta_b$ and $\beta_w$, respectively, and are
reported in the plot. Additional data for droplets without surfactant and with
surfactant concentrations below CAC are reported in
\blu{Figures~S3 and S4 of the SM}.
}
\end{figure}

\subsection{Bridge dynamics}
In the case $\theta_s > 90^\circ$, the pinching of the two droplets
takes place well above the substrate (Figure~\ref{fig:1}b), 
and only later does the bridge region make contact with the substrate (Figure~\ref{fig:1}c). 
In the case $\theta_s \leq 90^\circ$, the bridge starts to form on the substrate from the onset
of the coalescence process, which affects the bridge dynamics.
The rate of coalescence can generally be described by the
pace of the bridge growth in the direction normal as well
as parallel to the substrate. 
The size of the bridge {in these directions is $b$ and $w$, respectively,
both defined as shown in Figure~\ref{fig:1}.
Figure~\ref{fig:5} plots $b$ and $w$ above the CAC as a function of time
(see \blu{Figures~S3 and S4} in the SM for pure water and concentration below CAC.)
In the cases above CAC and for nonwettable substrates ($\theta_s > 90^\circ$), exponents
for $b$ are about 0.5, which are in line with the
case of freely suspended droplets.\cite{Arbabi2023b,Arbabi2023c} 
In contrast, in the case of wettable substrates, 
the exponent is higher reaching values of about 0.72,
which suggests a much faster dynamics in comparison
with the nonwettable substrates. Similarly,
the parameter $w$ shows a similar but weaker tendency for growth
with exponents of about 0.5 for both freely suspended
droplets\cite{Arbabi2023b,Arbabi2023c} and droplets on nonwettable substrates, up to about 0.65 for the wettable case. 
These higher values are agreeable with the 2/3 values seen for polymer droplets on wettable substrates.\cite{Arbabi2023}
In the case of water droplets, as a sanity check, exponents of $b$ as well as $w$ for
both wettable and nonwettable substrates are in
the range 0.60--0.66, which is in line with results
reported in the literature,\cite{eddi2013influence}
where a power law of 2/3 has been suggested 
(\blu{see Figure~S3a of the SM}).

When surfactant concentration is below the CAC
(\blu{Figure~S3b of the SM}), exponents for $b$ are in the
range 0.57--0.60 for non-wettable substrates (similar to pure water), but
a significant increase of the exponent is noted for
substrates reaching about 0.87, when the equilibrium contact angle of the droplet is
about $52^\circ$.
The exponents for $w$ are in the range 0.55--0.68, 
with the highest exponents observed in the 
case of equilibrium contact angles close to \blu{$70^\circ$}.

Finally, the initial thermal regime is well visible for
sessile droplets as in the case of freely suspended droplets\cite{Arbabi2023b,Arbabi2023c}
or sessile polymer droplets.\cite{Arbabi2023}
However, we notice both a decreasing extent of these fluctuations 
(smaller $b$, $w$) for sessile
droplets when $\theta_s < 90^\circ$, due to the additional
attraction and the contact with the substrate
which suppresses these fluctuations, but also concurrently a strong lengthening
of this regime in time, by almost an order of magnitude.

\begin{figure}[bt!]
\centering
\includegraphics[width=\columnwidth]{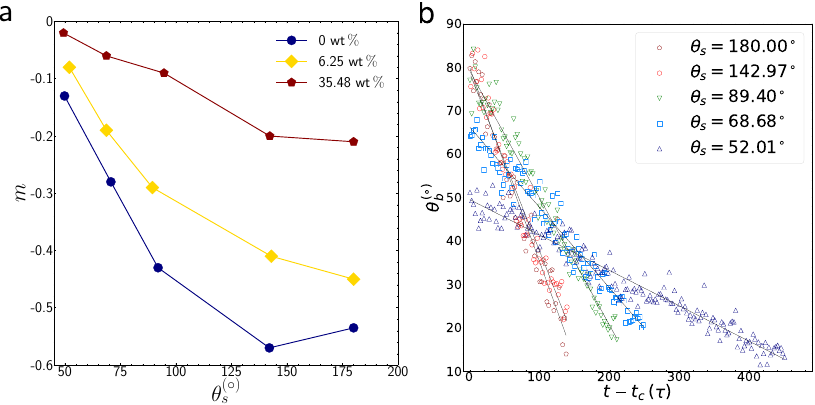}
\caption{\label{fig:6}
(a) Rate of change of angle $\theta_b$, $m=d\theta_b/d(t-t_c)$, as a function of the
equilibrium contact angle, $\theta_s$, that expresses
the wettability of the substrates with $\theta_s >90^\circ$ reflecting
non-wettable cases. $m$ is calculated from linear fits to the data of 
\blu{Figures~S5 and S6} of the SM, where a linear change of $\theta_b$ with time is observed
to a good approximation. The $\theta_s=180^{\circ}$ data is for freely suspended droplets
(no substrate). As an example here, panel b shows data for 
droplets with surfactant concentration 6.26~wt\%. 
}
\end{figure}

\subsection{Angle formed at the bridge}
We monitored the angle $\theta_b$ (Figure~\ref{fig:1}) as a function
of time for all cases by employing the method of Ref.~\citen{theodorakis2015modelling} \blu{(see Figure~S1 and text in the SM)}.
In particular, a layer parallel to the substrate at a distance $b$ was
considered, as indicated by the dashed line of Figure~\ref{fig:1}g.
The dimensions of the spherical-cap liquid-phase above this layer
were recorded in the $x$ and $y$ directions, as well as the distance
of the apex from the layer, which are used as input to calculate
$\theta_b$.\cite{theodorakis2015modelling} The approach
has been applied in both the right and the left parts of the merged droplets
and the average was taken, resulting in the final values reported for $\theta_b$.
Time traces of $\theta_b$ are plotted in \blu{Figures~S5 and S6} of the SM
for all cases considered here, while an example is shown in Figure~\ref{fig:6}.
We can observe that $\theta_b$ exhibits to a large degree
a linear behavior with time, which allows us to gather the slopes of the various curves
and thus monitor the rate of change (dynamics) of this angle, $m$, during coalescence.

These rates of change data are plotted in Figure~$\ref{fig:6}$.
The rate slows down as the substrate 
becomes more wettable, whether surfactant is present or not, but for
\blu{$\theta_s\gtrsim140^{\circ}$}, the freely suspended behavior 
is already reached. This is also directly seen in \blu{Figure~S5} of the SM. 
Greater surfactant concentration also slows down the process, as could be
suspected from the earlier suspended droplet studies \cite{Arbabi2023b,Arbabi2023c}, with pure water droplets being the fastest.
Hence, the use of surfactant facilitates the smoothing of the wedge formation 
at the bridge apex.

\begin{figure}[b]
\centering
\includegraphics[width=0.5\columnwidth]{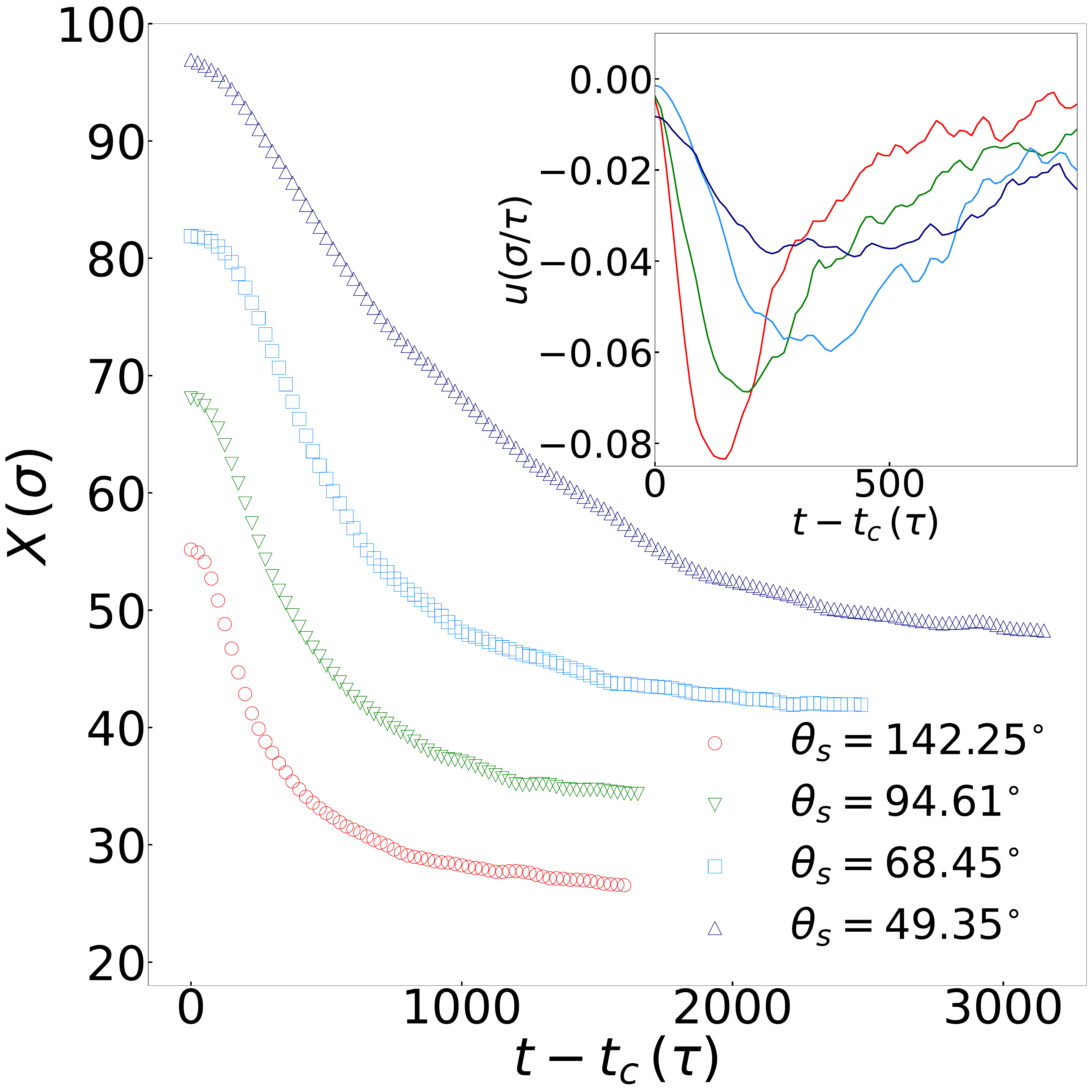}
\caption{\label{fig:7}
\blu{The distance between the centers of mass of the two droplets in the $x$-direction, $X$, (see Figure~\ref{fig:1}) }  as a function of time, $t$, starting from the
moment of first contact of the droplets, $t_c$, for surfactant-laden
droplets of concentrations above CAC (35.48~wt\%).
The inset shows the instantaneous velocity, $u=\dot{X}$.
Data for water droplets and surfactant-laden droplets with
concentration below CAC are shown in \blu{Figure~S7} of the SM.
}
\end{figure}

\subsection{Velocity of approach}
We calculated the coalescing system length, $X$, 
as defined in Figure~\ref{fig:1}, and
done in our previous work in the context of sessile polymer
droplets.\cite{Arbabi2023} Its value is approximately one droplet diameter.
Its time derivative, $u=\dot{X}$, provides a measure of how fast the droplets 
approach each other.  
Figure~\ref{fig:7} presents \blu{$X$ for droplets} above the CAC on 
various substrates with different wettability, while data for 
water droplets and droplets with surfactant with a concentration below CAC are given
in \blu{Figure~S7} of the SM. Overall, the velocity of approach is smaller in the case
of wettable substrates, irrespective of
surfactant concentration, paralleling what we saw with bridge sizes $b$ and $w$ 
and the angle formed at the bridge $\theta_b$. This is also true
for water droplets. The plots of \blu{Figure~S8} of the SM show that an
increase of surfactant concentration also significantly slows 
down the approach of the two droplets. 
This is mostly noticeable in the case of high concentrations, while
surfactant-laden droplets with concentration below CAC 
show a similar behavior to pure water.
Moreover, surfactant smooths the approach
of the two droplets, as can be seen by the change of shape of the
velocity of approach curves, particularly visible in \blu{Figure~S8} of the SM.
Finally, there is a clear shift in the moment of maximum velocity
to later times for more wettable substrates at higher surfactant concentration, 
though this shift is much smaller in the latter case.

\begin{figure}[bt!]
\centering
\includegraphics[width=\columnwidth]{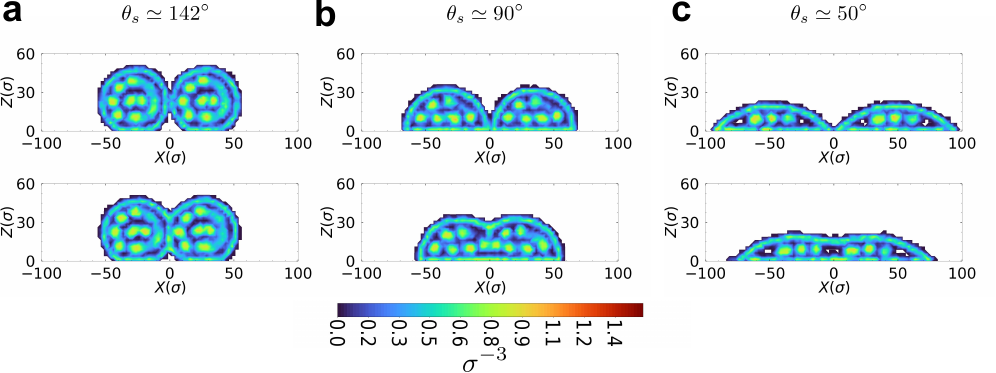}
\caption{\label{fig:8}
Surfactant density at the moment of pinching (upper panels)
and at a time when the bridge has been clearly developed 
(bottom panels) for above droplets with surfactant concentration
amounting to 35.48 wt\%. Here, cases for substrates with different
wettability are shown, namely,
(a) $\theta_s \simeq \blu{142}^\circ$, snapshots taken at $t_c$ and $t_c+100~\tau$, 
(b) $\theta_s \simeq 90^\circ$, snapshots taken at $t_c$ and $t_c+412.5~\tau$,
and (c) $\theta_s \simeq 50^\circ$, snapshots taken at
$t_c$ and $t_c+1352.5~\tau$.  
}
\end{figure}

\subsection{Density Profiles}
The density of surfactant molecules on a cross-section
parallel to the $x-z$ plane and passing through the center of mass of
the droplets is presented in Figure~\ref{fig:8}.
Also, \blu{Figure~S9} of the SM shows the corresponding density distribution of 
water molecules for various substrates. We observe
the formation of surfactant aggregates during the coalescence for substrates 
with equilibrium contact angles larger than $90^\circ$,
while it is absent for wettable substrates 
(see for example Fig.~\ref{fig:4}a--b and d--e versus g--h). 
Large concentrations of surfactant
are distributed at the LG and SL interfaces. Interior to this lies
an inner layer devoid of aggregates, 
which is particularly wide for the wettable droplets. 
Notably the wettable case has a much  smaller number of aggregates, 
despite the fact that all cases shown have
the same concentration. This is due to the larger area
of the LG and SL interfaces on wettable 
substrates, which might suggest that the CAC has a higher
value. The smaller thickness of the wettable droplets, also, 
hinders the accommodation of a larger number of surfactant aggregates.

\section{Conclusions}
\label{conclusions}
In this study, we have investigated the coalescence of sessile
surfactant-laden droplets on  substrates with different wettability, 
including those with equilibrium contact-angle 
higher as well as lower than $90^\circ$. We have explored the
influence of surfactant concentration both below and above the CAC,
and we juxtaposed our results with the case of pure water droplets.
In particular, we have elucidated the mass transport mechanism
in all these cases and explored the
dynamics of the coalescence process by following
the height and width of the bridge, 
the rate of change of the bridge angle, as well as the
velocity of approach of the droplets.

Overall, sessile droplets with $\theta_s \geq 90^\circ$
share similarities with freely suspended droplets, and for \blu{$\theta_s \simeq 140^{\circ}$}
already behave practically identically as if they were freely suspended.
In this case, the influence of the substrate on the coalescence process is
rather small. For example, the pinching region is
mainly driven by the interaction of the surfactant molecules
at the droplets' LG surface, as in the case of the freely
suspended droplets. In contrast, in the case of wettable
substrates ($\theta_s < 90^\circ$), we see that water
molecules are part of the pinching process,
a significant departure from the physics of the freely suspended case.
The mass transport of surfactant molecules during coalescence
also shows some differences between wettable and nonwettable
substrates, which mostly relates to their intensity.
A more notable difference is the absence of newly formed aggregates
as the bridge grows during coalescence in the case of wettable substrates.
This is due to the lower amount of surfactant at the pinching
region as the initial contact film has a far smaller area. 
In fact, the nature of the surface of first contact changes dramatically 
from a circular region to a thin contact line when the threshold of 
contact angle $\theta_s\approx90^{\circ}$ is crossed.
The smaller amount of space available in the bridge region between the
LG and SL interfaces during later evolution also contributes. 
The latter is related to a still open question of identifying changes in CAC
when the equilibrium contact angle of the droplets changes.

We also found that an increased wettability of 
the substrate leads to higher rate exponents for the 
growth of the bridge radius, $b$,
and its width, $w$, but generally exponents for $w$ are lower.
We confirm the approach to a $\approx\tfrac{2}{3}$ rate exponent at low 
contact angles suggested by several previous
studies,\cite{eddi2013influence,Arbabi2023} 
but see no evidence for the previously proposed linear 
growth.\cite{hernandez2012symmetric}
Overall, surfactant will decelerate the coalescence of the sessile droplets, 
when above CAC, 
but we observed higher exponents in the case of concentration below the CAC for
wettable substrates, which is not fully understood at the moment.
Similarly, the bridge angle, $\theta_b$,
changes at a faster pace in the case of water droplets and low concentration
and when the wettability of the substrate is lower (larger $\theta_s$). 
Finally, by analyzing the velocity of approach of the two droplets,
which is generally high at the initial stages of coalescence when the bridge
forms, we found that more wettable
substrates and a higher surfactant concentration will lead to smoother
changes in the velocity, less acceleration. 
We anticipate that our study provides fundamental insights into
the coalescence of sessile surfactant-laden droplets,
an important phenomenon that has previously mostly remained unexplored 
at the molecular-scale level.

\section{Supplementary Material}
\blu{The SM contains: details on the}
\blu{estimation of
the contact angle of a sessile droplet. Calibration of the} dependence of the contact angle of \blu{a} water droplet as a function of the droplet--substrate attraction parameter $\varepsilon_{ws}$. \blu{Table of the} water--substrate and surfactant--substrate interaction parameters. Further data and methodology details on the mass transport mechanism along \blu{with} the probabilities of surfactant moving between different areas of the droplets. \blu{Data on} the bridge angle, the velocity of approach, and the density profiles of the droplets, \blu{for a wider set of parameters than shown in the main text}.

\begin{acknowledgments}
This research has been supported by the 
National Science Centre, Poland, under
grant No.\ 2019/34/E/ST3/00232. 
We gratefully acknowledge Polish high-performance computing 
infrastructure PLGrid (HPC Centers: ACK Cyfronet AGH) for
providing computer facilities and support within
computational grant no. PLG/2022/015747.
\end{acknowledgments}

%\nocite{*}
%\bibliography{aipsamp}% Produces the bibliography via BibTeX.
%aipnum4-2.bst 2019-01-14 (MD) hand-edited version of apsrev4-1.bst
%Control: key (0)
%Control: author (8) initials jnrlst
%Control: editor formatted (1) identically to author
%Control: production of article title (0) allowed
%Control: page (1) range
%Control: year (1) truncated
%Control: production of eprint (0) enabled
\providecommand{\noopsort}[1]{}\providecommand{\singleletter}[1]{#1}%

\end{document}